\documentclass[aps,prl,twocolumn,showpacs,amsmath,amssymb,floatfix,groupedaddress]{revtex4}

\usepackage{graphicx}
\usepackage{bm}
\usepackage{color}

\begin{document}

\title{Contextual Values of Observables in Quantum Measurements}

\author{J. Dressel}
\author{S. Agarwal}
\author{A. N. Jordan}
\affiliation{Department of Physics and Astronomy, University of Rochester, Rochester, New York 14627, USA}

\date{\today}


\def\be{\begin{equation}}
\def\ee{\end{equation}}
\def\ba{\begin{align}}
\def\ea{\end{align}}
\def\la{\langle}
\def\ra{\rangle}
\def\ef{{\rm erf}}

\newcommand{\set}[1]{\mathfrak{#1}}              
\newcommand{\ob}[1]{\mathcal{#1}}                
\newcommand{\op}[1]{\hat{\bm #1}}                
\newcommand{\mat}[1]{ {\bm #1} }                 
\newcommand{\mean}[1]{\la#1\ra}                
\newcommand{\cmean}[2]{_{#1}\mean{#2}}           
\newcommand{\ket}[1]{\vert#1\ra}               
\newcommand{\bra}[1]{\la#1\vert}               
\newcommand{\ipr}[2]{\la#1\vert#2\ra}        
\newcommand{\opr}[2]{\ket{#1}\bra{#2}}           
\newcommand{\nrmsq}[1]{\ipr{#1}{#1}}             
\newcommand{\nrm}[1]{\sqrt{\nrmsq{#1}}}          
\newcommand{\pr}[1]{\opr{#1}{#1}}                
\newcommand{\Tr}[1]{\mbox{Tr}[#1]}    

\begin{abstract}
We introduce \emph{contextual values} as a generalization of the eigenvalues of an observable that takes into account both the system observable and a general measurement procedure.  This technique leads to a natural definition of a general conditioned average that converges uniquely to the quantum weak value in the minimal disturbance limit.  As such, we address the controversy in the literature regarding the theoretical consistency of the quantum weak value by providing a more general theoretical framework and giving several examples of how that framework relates to existing experimental and theoretical results.
\end{abstract}


\maketitle

The weak value (WV) of a quantum operator $\op{A}$ was introduced in 1988 by Aharonov, Albert, and Vaidman (AAV) \cite{AAV}.  They claimed that if a system is preselected in an initial quantum state $\ket{\psi_i}$, and post-selected on a final state $\ket{\psi_f}$, then the result of weakly measuring the operator $\op{A}$ was not its expectation value, but rather its \emph{weak value}:
\be
A_w = \bra{ \psi_f } \op{A} \ket{\psi_i} \, / \, \ipr{ \psi_f }{ \psi_i }.
\label{eq: AAV}
\ee
In the WV literature, it is customary to point out that for a bounded operator the WV can wildly exceed the eigenvalue range, and, unlike the expectation value, can be complex.  

The WV has been both a theoretically and experimentally fruitful concept because of these surprising features.   Theoretically, WVs have helped to understand counter-intuitive results that involve postselection, even resolving difficulties that arise in quantum mechanics such as Hardy's paradox and apparent superluminal travel \cite{paradox}.  They can be further seen as a single-system test of quantum mechanics:  whenever the WV exceeds the eigenvalue range, it rules out certain classes of hidden variable theories and is equivalent to violating a generalized Leggett-Garg inequality \cite{LG}.  On the experimental side, WVs that exceed the eigenvalue range have been observed in optical systems \cite{expts,WisemanPhoton}.  Indeed, WVs have recently found a practical application, because an anomalously large WV may be used to amplify a small shift in a system parameter.  This idea has been exploited in polarization \cite{kwiat} and interferometry \cite{howell} experiments to achieve subpicometer position resolution.  

In spite of this list of increasingly impressive accomplishments and insights, there is a long history of controversy regarding WVs \cite{controversy}.  Our primary concern here is not the ample philosophical discussion, but rather the physics controversy which we now detail.   (i) As defined, the WV (\ref{eq: AAV}) is generally complex, so how can it be measured?  This was artfully put by Landauer when he asked, ``Has anyone seen a stopwatch with complex numbers on its dial?'' \cite{quote}.  (Both the real and imaginary parts of the WV expression are physically relevant, but are measured with different experiments, see e.g. \cite{jozsa}.)   (ii) Another objection is that the AAV formula (\ref{eq: AAV}) is not unique, and that other results for the WV can arise with a slightly different measurement setup \cite{Parrott}.  (iii) The AAV formula (\ref{eq: AAV}) only applies to pure states and arbitrarily weak measurements, so is there a generalization to the cases of mixed states and finite strength measurements?  (iv) Perhaps most worrisome, if we view the WV as a weighted average of the eigenvalues of the operator, and the WV exceeds the eigenvalue range, then we can immediately conclude that there must be negative probabilities weighting the eigenvalues.  This is dramatically illustrated in the three-box problem \cite{controversy}.  Taken together, these objections are a formidable challenge to the theoretical consistency of the concept of the WV.

The purpose of this Letter is to take the thesis of the second paragraph together with the antithesis of the third and to develop a synthesis.  This will be accomplished by going beyond thinking about an observable in terms of its eigenvalues and advancing the concept that the measurement results must be interpreted within their own context.  This idea will give rise to the notion of the \emph{contextual values} (CV) of an observable.  The synthesis occurs by showing how the WV can be subsumed as a special case in the CV formalism.  This formalism contains additional information that is able to clarify the objections listed above in an organic fashion.  In addition, we will derive a generalized WV formula that can be practically applied to a much larger class of problems that involve postselected averages.  We also provide illustrative examples, and demonstrate that we can recover a number of predictions for postselected averages in the literature that had been calculated for specific situations using other, more involved techniques.

\emph{Contextual values}.---The idea behind CV stems from the observation that the intrinsically \emph{measurable} quantities in the quantum theory are the outcome probabilities for a particular measurement setup.  In order to calculate averages of an observable from the measured probabilities, it is empirically necessary to assign values to each outcome as a separate step.  This secondary step becomes implicit in the operator formalism, so is rarely mentioned explicitly.

Under projective measurements such a value assignment seems trivial: the outcomes are perfectly distinguishable and easily assigned meaningful values, which become the eigenvalues for the observable operator.  However, when generalizing the measurement to nonprojective outcomes, one must be careful to correctly match the measured probabilities to a consistent value set.  We will now show that the observable operator can be expanded in multiple ways corresponding to the possible measurement strategies and value assignments (\ref{eq: Contextual Decomposition}).

To illustrate how the observable values change, we consider a general measurement that is fully characterized by a set of $N$ measurement operators, $\set{M} = \{\op{M}_j\}$, which represent the operation of an $N$-outcome measurement apparatus on the system.  This \emph{measurement context} generates a corresponding positive operator-valued measure (POVM), $\set{E} = \{\op{E}_j = \op{M}^\dagger_j \op{M}_j\}$, which, in conjunction with an arbitrary density operator $\op{\rho}$, determines a probability measure, $P_j = \Tr{\op{E}_j \op{\rho}}$, on the space of outcomes.  Empirically, these probabilities can be measured as the relative frequencies of obtaining each outcome.

Suppose that we wish to \emph{reconstruct} the average value of an observable $\ob{A}$ using this setup.  In order to obtain the correct average as a weighted sum,
\be
  \mean{\ob{A}} = \sum_j \alpha_j P_i = \sum_j \alpha_j \Tr{\op{E}_j \op{\rho}},
  \label{eq: POVM Average}
\ee
we must explicitly assign a set of values $\{\alpha_j\}$ to the $N$ outcomes of the apparatus.  

We require that such an average be independent of $\set{M}$, and thus a property of the state $\op{\rho}$ itself, which constrains the possible value assignments.  Specifically, (\ref{eq: POVM Average}) should be equivalent to the average under a projective context, $\set{P} = \{\op{\Pi}_k\}$,
\be
  \mean{\ob{A}} = \sum_k a_k \Tr{\op{\Pi}_k \op{\rho}} = \Tr{ \op{A} \op{\rho} },
  \label{eq: PVM Average}
\ee
where we have identified the projection-valued measure (PVM) and the corresponding eigenvalues $\{a_k\}$ as the spectral decomposition of a Hermitian operator, $\op{A}$.

We now assert that the values $\{\alpha_j\}$ should be determined solely by $\set{M}$ so the equality between (\ref{eq: POVM Average}) and (\ref{eq: PVM Average}) should hold for any state $\op{\rho}$. It follows that the operator $\op{A}$ can be expanded in multiple ways corresponding to possible contexts,
\be
  \label{eq: Contextual Decomposition}
  \op{A} = \sum_j \alpha_j \op{E}_j = \sum_k a_k \op{\Pi}_k.
\ee
This operator identity defines the \emph{contextual values} $\{\alpha_j\}$ of an operator $\op{A}$ under a compatible measurement context $\set{M}$ as a generalization of its eigenvalues that allows the full empirical reconstruction of the associated physical observable $\ob{A}$.  

Indeed we can use this operator equality to obtain the $n$\textsuperscript{th} moment of $\ob{A}$ using the same experimental setup,
\be
  \label{eq: Moment Reconstruction}
  \mean{\ob{A}^n} = \sum_{j_1,\ldots,j_n} \alpha_{j_1}\ldots\alpha_{j_n} \Tr{\op{E}_{j_1}\ldots\op{E}_{j_n}\op{\rho}},
\ee
provided that the $N$ measurement operators in $\set{M}$ and $\op{A}$ all commute. In that case the correlation probabilities, $P_{j_1,\cdots,j_n} = \Tr{\op{M}_{j_n}\cdots\op{M}_{j_1}\op{\rho}\op{M}^\dagger_{j_1}\cdots\op{M}^\dagger_{j_n}}$, appearing in (\ref{eq: Moment Reconstruction}) can be measured as the relative frequencies of the $N^n$ outcome \emph{sequences} of $n$ repeated measurements.

In addition to permitting the reconstruction of the moments of the observable, its CV are of great interest in their own right.  Note that the moments of the CV themselves, $\sum_j \alpha_j^n P_i$, only coincide with the moments of $\ob{A}$ for $n=1$.  Such moments nevertheless contain important physics about both the context $\set{M}$ and the observable $\ob{A}$, as we consider shortly. 

The construction (\ref{eq: Contextual Decomposition}), in general, will fail if $N < \mbox{dim}\op{A}$, will be unique if $N = \mbox{dim}\op{A}$, and will be underspecified if $N > \mbox{dim}\op{A}$.  The latter case results in an infinite number of possible solutions, $\{\alpha_j\}$.  As such, we propose that the physically sensible choice of CV is the least redundant set uniquely related to the eigenvalues through the Moore-Penrose pseudoinverse. 

To illustrate the construction of the least redundant set of CV, we consider the case when $\{\op{M}_j\}$ and $\op{A}$ all commute.  It follows that they can all be diagonalized in the same PVM, and (\ref{eq: Contextual Decomposition}) is isomorphic to a matrix equation $\vec{a} = \mat{F}\vec{\alpha} = (\sum_j F_{kj} \alpha_j)$, with the components of $\mat{F}$ given by, $F_{kj} = \Tr{\op{\Pi}_k \op{E}_j}$.  To find the pseudoinverse of the matrix, $\mat{F}$, we find its singular value decomposition, $\mat{F} = \mat{U} \mat{\Sigma} \mat{V}^T$, where the orthogonal matrices, $\mat{U}$ and $\mat{V}$, are composed of the eigenvectors of $\mat{F}\mat{F}^T$ and $\mat{F}^T \mat{F}$, respectively, and $\mat{\Sigma}$ is a diagonal matrix composed of the singular values of $\mat{F}$ \cite{inversion}.  The pseudoinverse of $\mat{F}$ can be constructed as, $\mat{F}^{+} = \mat{V} \mat{\Sigma}^{+} \mat{U}^T$, where $\mat{\Sigma}^{+}$ is the diagonal matrix constructed from $\mat{\Sigma}^T$ by inverting all nonzero singular values.  It then follows that a uniquely specified set of CV can be defined as $\vec{\alpha}_0 = \mat{F}^{+} \vec{a}$.  Other solutions of (\ref{eq: Contextual Decomposition}) can be written $\vec{\alpha} = \vec{\alpha}_0 + \vec{x}$, where $\vec{x}$ is in the null space of $\mat{F}$.

\emph{Conditioned averages}.---Suppose we now wish to generalize the notion of an observable average by conditioning the average on the result of a second measurement.  To do so we weight the CV, $\{\alpha^{(1)}_j\}$, already determined by the first measurement context, $\set{M}^{(1)}$, with a set of conditional probabilities generated from a second measurement.

A second measurement is performed on the system with a context, $\set{M}^{(2)} = \{\op{M}^{(2)}_f\}$, yielding the full context for the (ordered) sequential measurements, $\set{M}^{(1,2)} = \{\op{M}^{(2)}_f \op{M}^{(1)}_j \}$.  The corresponding POVM, $\set{E}^{(1,2)} = \{ \op{E}^{(1,2)}_{jf} = \op{M}^{(1)\dagger}_j \op{M}^{(2)\dagger}_f \op{M}^{(2)}_f \op{M}^{(1)}_j \}$, gives the measurable probabilities, $P_{jf} = \Tr{\op{E}^{(1,2)}_{jf} \op{\rho}}$ and $P_f = \sum_j P_{jf}$.  From these probabilities we can define a conditional probability in the usual way, $P_{j|f} = P_{jf}/P_f$, and thus define our main result, the \emph{conditioned average} of an observable,
\be
  \cmean{f}{\ob{A}} = \sum_j \alpha^{(1)}_j P_{j|f} = \frac{\sum_j \alpha^{(1)}_j \Tr{\op{E}^{(1,2)}_{jf}\op{\rho}}}{\sum_j \Tr{\op{E}^{(1,2)}_{jf} \op{\rho}}}.
  \label{eq: Conditioned Average}
\ee
Unlike $\mean{\ob{A}}$, $\cmean{f}{\ob{A}}$ is explicitly dependent on $\set{M}^{(1)}$ and the CV.  Hence, it must generally be interpreted as encoding information not just about the observable $\ob{A}$, but also about the measurement context itself.

Conditioned averages must be bounded by the minimum and maximum CV.  However, since the CV range is not the same as the eigenvalue range for the observable, values of $\cmean{f}{\ob{A}}$ outside the eigenvalue range may be obtained.

\emph{Weak values}.---To find the weak limit of (\ref{eq: Conditioned Average}) we note that any measurement context continuously connected to the identity operation can be decomposed into the form $\set{M} = \{\op{U}_j(g)\op{E}^{1/2}_j(g)\}$, where $g$ is a measurement strength parameter, and $\op{U}_j(g) = \exp[i g \op{G}_j]$ are unitary operators generated by Hermitian operators $\op{G}_j$ by Stone's theorem.  Since the POVM elements are bounded and partition unity, they can also be expanded as $\op{E}_j(g) = p_j \op{1} + g \op{E}'_j + \mathcal{O}(g^2)$, where $\sum_j p_j = 1$.

Writing the initial context in (\ref{eq: Conditioned Average}) to first order in $g$, we find that as $g\to0$, the weak limit generally depends explicitly on $\{\op{G}_j\}$ and $\{\alpha_j\}$, and thus will change depending on how it is measured and how the CV are chosen (see also \cite{Parrott}).  However, if $\forall j, [\op{G}_j,\op{\rho}] = 0$, so the state is minimally disturbed, then the context-dependence vanishes and a generalized WV \cite{Wiseman2002} is \emph{uniquely} defined as the quantity
\be
  A_w = \Tr{\op{E}^{(2)}_f \{\op{A},\op{\rho}\}} \, / \, 2 \Tr{\op{E}^{(2)}_f \op{\rho}},
  \label{eq: Weak Value}
\ee
where $\{\cdot,\cdot\}$ denotes the anticommutator.  For a pure initial state, $\op{\rho} = \pr{\psi_i}$, a ``pure POVM,'' $\forall j, \op{U}_j = \op{1}$, and a strong final measurement, $\op{E}^{(2)}_f = \pr{\psi_f}$ (as considered by AAV and most subsequent applications), (\ref{eq: Weak Value}) can be written as the \emph{real part} of (\ref{eq: AAV}) \cite{strong cond av}.  

\emph{Photon polarization}.---As a first example, a photon polarization measurement of tunable strength is detailed in \cite{WisemanPhoton} and can be readily interpreted using CV.  The Stokes observable $\ob{S}_z$ is measured in the horizontal-vertical polarization basis with the context $\op{M}_+ = \gamma \op{\Pi}_H + \bar{\gamma} \op{\Pi}_V$ and $\op{M}_- = \bar{\gamma} \op{\Pi}_H + \gamma \op{\Pi}_V$, leading to the POVM, $\op{E}_{\pm} = (1/2)\left(\op{1} \pm g \op{\sigma}_z\right)$, where $g = \gamma^2 - \bar{\gamma}^2$.  The CV for this scenario are uniquely $\pm 1/g$, which exceed the eigenvalue range for $|g| < 1$.

Computing the conditioned average (\ref{eq: Conditioned Average}) of $\ob{S}_z$ with an initial state of $\ket{\psi} = \alpha \ket{H} + \beta \ket{V}$ and a final state $\ket{f} = \left(\ket{H} - \ket{V}\right)/\sqrt{2}$ yields
\be
  \cmean{f}{\ob{S}_z} = (|\alpha|^2 - |\beta|^2) \, / \, (1 - 4\gamma \bar{\gamma} \mbox{Re}[\alpha \beta^*]),
\ee
which is the result obtained in \cite{WisemanPhoton} through direct computation using an ancilla system.

\begin{figure}[t]
\begin{center}
\leavevmode 
\includegraphics[width=8.5cm]{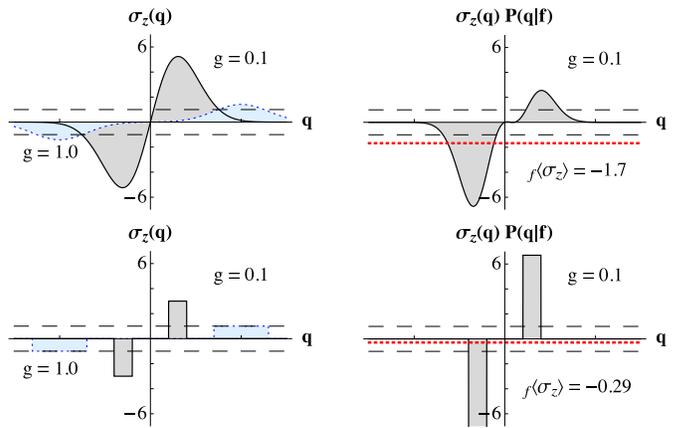} 
\caption{(color online) CV (left) and conditioned CV (right) computed using the setup described above Eq. (\ref{eq: AAV Weak Value}) with a coupling strength $g = 0.1$ and a pointer distribution width $\sigma = 0.3$.  The preparation state is $\psi(47\pi/32)$, and postselection state is $f(\pi/2)$.  Two detector distributions are compared: Gaussian (top) with conditioned average (right, dotted line) violating the eigenvalue range (dashed lines), and box (bottom) showing the effect of the shape on the WV convergence rate.  Strong measurement regime CV for each distribution with $g = 1.0$ are also shown for comparison (left, dotted line).} 
\label{fig1}
\end{center}
\vspace{-6mm}
\end{figure}

\emph{AAV setup}.---For the original AAV/von-Neumann setup \cite{AAV}, implemented using photons in \cite{expts}, an interaction Hamiltonian $\op{H}_{\mbox{int}} = g \delta(t - t_0) \op{\sigma}_z \otimes \op{p}_D$ of a qubit with a detector having momentum operator $\op{p}_D$ leads to a continuous set of measurement operators on the system $\op{M}(q) = \ipr{q - g\op{\sigma}_z}{\Phi_D}$, where $\ket{\Phi_D}$ is the initial detector state, $q$ is the position of the detector pointer, and $g$ is the coupling strength.  Hence the POVM is simply the initial detector probability density shifted in position by an operator, $\op{E}(q) = P_D(q - g\op{\sigma}_z) = |\ipr{q - g\op{\sigma}_z}{\Phi_D}|^2$.

Performing the pseudoinversion \cite{inversion} leads to the least redundant set of CV, $\sigma_z(q)$, for the $z$-component of the pseudospin,
\begin{align}
  \label{eq: AAV Contextual Values}
  &\: \sigma_z(q) = \left[ P_D(q - g) - P_D(q + g) \right] \, / \, \left( a - b(g) \right), \\
  &\: a = \int P_D(q)^2 \, dq; \: b(g) = \int P_D(q - g) P_D(q + g) \, dq. \nonumber
\end{align}
From these expressions it is immediately clear that if the shift $g$ is small compared to the width of the initial distribution $P_D(q)$, then the POVM will be nearly proportional to the identity and the measure of overlap, $b(g)$, will be nearly equal to the distribution constant $a$.  Hence, the measurement will be weak and the CV will diverge independently of any specific system states.  

For a Gaussian distribution, $P_D(q)=\exp[-q^2 / 2\sigma^2]/\sigma\sqrt{2\pi}$, $a=1/2\sigma \sqrt{\pi}$ and $b(g)=a\exp[-(g/\sigma)^2]$.  Hence, the CV are $\sigma_z(q) = \sqrt{2} \exp[-q^2/2\sigma^2]\left[ \sinh(q g/\sigma^2) / \sinh(g^2/2\sigma^2) \right]$ \cite{aav q/g}.  With an initial state $\psi(\alpha) = \left(\cos(\alpha/2),\sin(\alpha/2)\right)$ in the $z$ basis conditioned on a final state $f(\pi/2) = (2)^{-1/2}\left(1,1\right)$, (\ref{eq: Conditioned Average}) yields
\be
  \cmean{f}{\ob{S}_z} = \cos\alpha \, / \, (1 + \sin\alpha\exp[-g^2/2\sigma^2]).
  \label{eq: AAV Weak Value}
\ee
The strong limit as $(g/\sigma)\to\infty$ is $\cos\alpha$ and the weak limit as $(g/\sigma)\to0$ is the WV $\cot(\alpha/2 + \pi/4)$.  The same WV is also obtained for any $P_D(q)$ according to (\ref{eq: Weak Value}), though the speed of convergence depends on the distribution shape.  Figure \ref{fig1} illustrates the effect of the shape by comparing Gaussian and box distributions.

\emph{Quantum point contact (QPC) detector}.---A generalization of the AAV result to continuous measurements in solid state detectors is given in \cite{LG}.  The measurement operators are constructed in terms of an average current through a QPC, $\bar{I} = (1/t)\int_0^t I(t')\, dt'$, and two characteristic currents, $I_1$ and $I_2$ that indicate the position of an electron in a nearby double quantum dot.  The current detection probabilities are assumed Gaussian with a variance of $\sigma^2 = S_I / 2t$, where $S_I$ is the detector shot noise power and $t$ is the averaging time for the current.

\begin{figure}[t]
\begin{center}
\leavevmode \includegraphics[width=6.5cm]{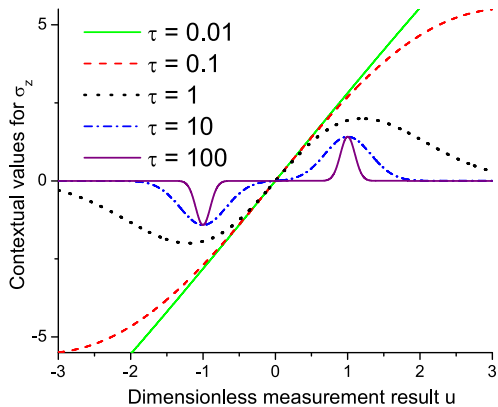} 
\caption{(color online) QPC CV for a range of measurement times and a fixed coupling constant.  For small times $\sigma_z(u) \to 2\sqrt{2} u$, while for long times $\sigma_z(u)$ is only nonzero at $\pm 1$.}
\label{fig2}
\end{center}
\vspace{-6mm}
\end{figure}

The system can be mapped to the AAV case by making the identifications $I_0 = (I_1 + I_2)/2$, $q = \bar{I} - I_0$, $g = (I_1 - I_2)/2$, and $\sigma^2 = S_I / 2t$.  Since the characteristic currents for the electron position are fixed, the coupling strength $g$ is also fixed.  Hence it is useful to consider instead the dimensionless variables, $u = q/g$, and $\tau = (g/\sigma)^2 = t/T_m$, where $T_m$ is the characteristic measurement time.  Figure \ref{fig2} shows a range of CV for varying $\tau$.

The system is started with an initially arbitrary density matrix $\op{\rho} = [\rho_{ij}]$ written in the $z$ basis.  After being measured for the averaging time $t$, it is then rotated around the $\op{\sigma}_x$ axis by an angle $\theta$ and measured strongly to be in a final state $\ket{f} = \ket{+1}$ along the new $z$ axis.  Computing the conditioned average using (\ref{eq: Conditioned Average}) yields
\be
  \cmean{f}{\ob{S}_z} = \frac{\cos^2(\frac{\theta}{2}) \rho_{11} - \sin^2(\frac{\theta}{2}) \rho_{22}}{\cos^2(\frac{\theta}{2}) \rho_{11} + \sin^2(\frac{\theta}{2}) \rho_{22} - \sin\theta\mbox{Im}[\rho_{12}]e^{-\tau/2}},
\ee
which matches the result in \cite{LG} computed using the quantum Bayesian approach \cite{aav q/g}.  

\emph{Conclusions}.---We have shown how to completely reconstruct an observable using an arbitrary measurement setup by defining \emph{contextual values} (\ref{eq: Contextual Decomposition}), or generalized eigenvalues, that account for the measurement context.  The approach allows a conditioned average (\ref{eq: Conditioned Average}) to be defined that correctly reproduces existing results in the literature and converges uniquely to a generalized weak value (\ref{eq: Weak Value}) in the minimum disturbance limit.

This conditioned average addresses much of the controversy surrounding weak values in the literature, since it (i) is a purely real empirical average, (ii) converges to the real part of the standard weak value (\ref{eq: AAV}) uniquely in the minimum disturbance limit with pure initial and final states, (iii) applies to any measurement strength, arbitrary quantum density operators, and POVM postselections, and, (iv) is constructed from well-behaved, measurable probabilities.

\begin{acknowledgments}
This work was supported by the NSF Grant No. DMR-0844899, ARO Grant No. W911NF-09-1-0417, and a DARPA DSO Slow Light grant.
\end{acknowledgments}


\begin{thebibliography}{99}
\bibitem{inversion}
The singular value decomposition can be readily generalized to an operator $\op{A}$ and/or POVM with infinite dimension as long as the resulting operator $\op{F}$ is compact.
\bibitem{aav q/g}
If $P_D(q)$ is symmetric about the mean, $\cmean{f}{\ob{S}_z}$ becomes independent of the choice of CV.  Hence, the generic CV often implied in the literature, $\sigma_z(q) = q/g$, give the same result as the least redundant CV.
\end{thebibliography}
\end{document}